\documentclass[prr,onecolumn,superscriptaddress]{revtex4}

\usepackage{amssymb,amsbsy}
\usepackage{graphicx}
\usepackage{rotating}
\usepackage{dcolumn}
\usepackage{amsmath}
\usepackage[abs]{overpic}

\newcommand{\abs}[1]{\vert #1 \vert}

\begin{document}
\title{Online geolocalized emotion across US cities during the COVID crisis: Universality, policy response, and connection with local mobility}

\author{Shihui Feng}
\affiliation{Unit of Human Communication, Development, and Information Sciences, Faculty of Education, The University of Hong Kong, Hong Kong, China}

\author{Alec Kirkley}
\affiliation{Department of Physics, University of Michigan, Ann Arbor, Michigan, USA}

\begin{abstract}
As the COVID-19 pandemic began to sweep across the US it elicited a wide spectrum of responses, both online and offline, across the population. To aid the development of effective spatially targeted interventions in the midst of this turmoil, it is important to understand the geolocalization of these online emotional responses, as well as their association with offline behavioral responses. Here, we analyze around 13 million geotagged tweets in 49 cities across the US from the first few months of the pandemic to assess regional dependence in online sentiments with respect to a few major topics, and how these sentiments correlate with policy development and human mobility. Surprisingly, we observe universal trends in overall and topic-based sentiments across cities over the time period studied, with variability primarily seen only in the immediate impact of federal guidelines and local lockdown policies. We also find that these local sentiments are highly correlated with and predictive of city-level mobility, while the correlations between sentiments and local cases and deaths are relatively weak. Our findings point to widespread commonalities in the online public emotional responses to COVID across the US, both temporally and relative to offline indicators, in contrast with the high variability seen in early local containment policies. This study also provides new insights into the use of social media data in crisis management by integrating offline data to gain an in-depth understanding of public emotional responses, policy development, and local mobility. 
\end{abstract}
\maketitle

\section{Introduction}
When an unpredictable adverse event strikes, it is critical to understand public emotional and behavioral responses in order to support policy development and relief management. Given its prevalent use during crises and strengths for facilitating connectivity and collective efforts among individuals, social media has been used increasingly as an effective digital source for assessing collective responses to extreme crises within and beyond affected communities \cite{gao2011harnessing,neubaum2014psychosocial,li2017reasoning}. However, an inherent challenge for integrating social media usage into crisis management is that most social-media based findings can help us understand the perspectives of online users regarding a crisis, but have limited capability to contribute to the development of on-site response strategies and relief activities. Bridging the gap between the theoretical significance and practical applications of social media usage in crisis management requires us to connect online and offline information in order to examine the relationships between online emotional responses and offline events and behavioral responses during crises. Geolocalized information on social media is the key to match online user-generated information with offline local situations. In this study, we analyze the online geolocalized emotion (OGE)---the sentiments derived from a set of geotagged content on social media---towards COVID-19 using Twitter data. The relationships of OGE with COVID-related polices and offline mobility are also assessed, in an attempt to uncover the interdependence among public emotional responses, policy development, and local mobility during this significant public health crisis. \\

COVID-19, an infectious disease caused by a novel coronavirus, had infected around 3.2 million people around the world as of May 1st, the end of the time period analyzed in this study, and has infected over 30 million people to date. This long-lasting and highly contagious epidemic has impacted every aspect of society, and is considered by many the greatest global challenge since World War II. In response to this crisis, beyond the enormous health care effort, scientific communities in various disciplines have been working together to better understand the economic, societal, and social effects of this crisis. As part of this effort, a number of studies have been conducted to examine the use of social media (e.g. Twitter, Weibo, and Facebook) during COVID-19. The topics of these studies mainly fall into two categories: assessing public mental health and feelings such as anxiety and fear towards COVID-19 \cite{gao2020mental,roy2020study,zhang2020impact,barkur2020sentiment,mckay2020anxiety,zhang2020monitoring,gupta2020COVID}, and diffusion of crisis-relevant and false information \cite{sun2020early,pulido2020COVID,tao2020nature,thelwall2020retweeting,memon2020characterizing,sharma2020COVID}. A few studies have utilized geotagged data on social media to map and predict the number of infected cases \cite{shen2020using} and offline mobility \cite{porcher2020social,huang2020twitter}. 
For instance, Huang et al \cite{huang2020twitter} propose an approach to capture offline mobility in certain geographic regions through online geotagged data, in order to assess responsiveness to protection measures. In this study, we aim to examine geolocalized emotional responses toward COVID and explore their relationships with national and local policy development as well as offline human mobility. The closest work we find to our study is that by Porcher and Renault \cite{porcher2020social}, who analyzed the relationship between the number of tweets relevant to social distancing and trends in human mobility at a state-level in the US using 402,005 tweets, finding through OLS regressions that an increase in online discussion about social distancing is associated with a decrease in mobility with a one-day lag. Our study contributes to this line of discussion with a much larger dataset, examining online responses with geolocalized sentiments (rather than tweet counts) and using more comprehensive statistical methods to analyze the relationships between time series data to assess the connections between online responses and offline mobility, as well as the policy effects on OGE dynamics. \\

The research questions leading this study are: 1) What are the characteristics of OGE dynamics towards COVID-19 across US cities?; 2) How do federal and local policies affect OGE?; 3) What are the relationships of OGE with offline mobility and infection data? Around 13 million geotagged tweets and daily city-level mobility data from February 26th to May 1st 2020 are used in this study to address these questions. Using measures derived from tweet sentiments, we quantify daily geolocalized sentiment regarding five key COVID-related subtopics across different locations, as well as its coherence across these locations, finding strong universal trends across cities and high correlations at the daily level. We also find through a rigorous intervention analysis that, despite the high similarity in temporal trends, cities can be differentiated in their OGE dynamics through their responses to federal and local policy decisions. Finally, we examine the impact of online geolocalized emotion on offline factors, including daily cases, deaths and mobility measures. We observe high correlations between mobility measures and OGE, while the association between the epidemic measures and OGE is relatively weak or absent in most cities. We further analyze the statistical connection between OGE dynamics and mobility data by assessing the theoretical predictability of the mobility measures from past OGE data using a Granger causality framework, finding significant causality in most cities with respect to many OGE subtopics. This study provides an empirical demonstration of an effective analytic framework integrating social media data with offline data to examine the relationships between public emotional responses, policy development, and local mobility during crises. The findings of this study can help us gain a holistic understanding of the relationships between public emotional and behavioral responses in the US during COVID-19, as well as strengthen the practical significance of social media analytics for supporting the development of response strategies and activities. 

\section{Methods}
\subsection{Data}
To efficiently isolate COVID-related tweets with geotagged data, we identify the subset of tweets collected in \cite{qazi2020geocov19} that have `geo' or `user\_location' attributes (as opposed to the inferred locations identified by the study) within the 100 Metropolitan Statistical Areas (MSA's) with the highest populations \cite{censusMSA}. Geolocations with `city' labels corresponding to subregions within each MSA were mapped to their associated MSA, and `geo' attributes were given priority over `user\_location' attributes if both were available. Tweets were dated over the period from February 1st 2020 to May 1st 2020, and to reduce statistical noise, we only study MSA's with an average of $\geq 500$ tweets per day over this period, which reduces the dataset to 49 cities. After initial inspection, we reduced the timeframe of study to begin February 26th, as all cities studied had a significant jump in tweet count on this day, with over 100,000 total daily tweets for the first time. This date also corresponds with the day when the CDC confirmed the first community transmission within the US, and so it is a key date in the evolution of the disease spread in the US. The final dataset consists of $12,670,890$ tweets across the 49 cities, with a minimum count of $44,575$ tweets (Grand Rapids, MI), a maximum count of $1,551,182$ tweets (Washington, DC), and a median count of $170,234$ tweets. The specific cities studied can be seen in Figure 3.\\

Despite its high volume, Twitter data produces inherently noisy estimates in sentiment classification analyses due to its sparsity and high composition of non-standard characters \cite{saif2012alleviating,srivastava2017challenges}. To mitigate these issues as much as possible, we limit our analyses to aggregate trends in the polarity of tweets. Tweet sentiment was analyzed using the Amazon Comprehend API \url{https://aws.amazon.com/comprehend/}, which has been shown to outperform other off-the-shelf methods for correctly identifying tweets with positive or neutral sentiment, and vastly outperforms more naive methods \cite{carvalho2020off}. In a period characterized by excess negative sentiment \cite{lwin2020global}, the ability to correctly identify tweets with positive and neutral sentiment is of the utmost importance, and so we opt for the Comprehend API due to this strength. The API returns confidence scores (normalized to sum to $1$) associated with the sentiment classifications \{Positive, Neutral, Negative, Mixed\} for the tweet being analyzed. However, as the method to obtain these scores is proprietary and confidential, we choose to simply utilize the sentiment of highest confidence as the classification for each tweet to maintain a simple interpretation of the measures we discuss. (Initial tests revealed that considering the confidence scores in weighted variants of our measures made little to no qualitative difference anyway.) The final dataset, consisting of Tweet ID (in compliance with the Twitter terms of use agreement) and primary sentiment classification for all $12,670,890$ tweets is available at \url{https://github.com/aleckirkley/US-COVID-tweets-with-sentiments-and-geolocations/}.\\

For mobility data, we collect the daily city-level values for driving and walking from the Apple mobility trends reports \url{https://www.apple.com/COVID19/mobility}, which give the relative volume of Apple Maps route requests per city compared to a baseline volume on January 13th, 2020. Cities in this dataset are delineated by their corresponding greater metropolitan area, and so are geographically bounded to the same regions as the tweet data. We also utilize COVID case and death time series data in Fig. 3 to compare the correlations between sentiments and these attributes with the correlations we see between the sentiments and mobility measures. The values for confirmed daily cases and deaths in all counties within each MSA were aggregated from the JHU CSSE repository \url{https://github.com/CSSEGISandData/COVID-19}.\\

In order to assess how the dynamics of online geolocalized emotion were affected by major policy events in the early stages of the epidemic in the US, we identify three federal policies and one local policy to use as reference policy events. Due to the generally decentralized and casual approach to containment through policy interventions taken by the federal government during the early stages of the epidemic in the US \cite{carter2020making,parodi2020containment,moon2020fighting}, it is difficult to clearly isolate key federal policy actions taken during this period. We thus identify the dates associated with three policy announcements during the time period studied that may reflect the general opinion about the state of the epidemic from the viewpoint of the federal government: (1) March 13, the date the US declared COVID-19 a national emergency; (2) March 29, the date President Donald Trump officially extended social distancing guidelines---discouraging nonessential workplace attendance and travel, eating at restaurants, and gatherings of more than ten people---through the end of April; (3) April 16, the date President Trump released a set of guidelines to states for reopening, based on the condition of the individual state. For local policy, we identify the dates each city instituted a shelter in place order, based on the dates of these policy announcements at the state level. We also record the dates that cities ended their local shelter in place orders (again based on state policies), if this occurred within the timeframe studied, and these are accounted for in the intervention analysis in Fig. 2.     

\subsection{Online geolocalized emotion measures}
Two measures are proposed in this section to analyze the online geolocalized emotion (OGE) in the 49 cities at micro- and macro-levels. The first is an intuitive measure used to analyze the average polarity of daily online sentiments in each city, which quantifies the positive or negative tendency of online sentiments toward COVID-19 in each geolocation group. The second measure analyzes the coherence of online sentiments across the 49 cities, and is used to examine the uniformity of online emotional responses at the country-level across time with respect to multiple subtopics. As the effects of COVID-19 are multifaceted, in addition to understanding the online emotional response based on all COVID-relevant geotagged content, people's opinions about five important subtopics are also assessed: the Trump administration (abbr. ``TA"), China (abbr. ``China"), social distancing/quarantining (abbr. ``distancing"), face masking (abbr. ``mask"), and the economy (abbr. ``economy"). We also denote tweets not restricted to any particular subtopic using the abbreviation ``overall". With the subtopics of interest chosen ahead of time, to identify keywords for each of these topics we look at the number of tweets related to all unique words in the full set of tweets, and identify high-frequency keyword sub-strings associated with each of the five topics. The keyword sub-strings identified with each topic are (all in lowercase as were the cleaned tweets):
\begin{itemize}
    \item ``TA": \{``trump",``pence"\} 
    \item ``China": \{``china",``chinese",``wuhan"\}
    \item ``distancing": \{``quarantin",``lockdown",``social distanc"\}
    \item ``mask": \{``mask",``ppe"\}
    \item ``economy": \{``econom",``stock market",``dow",``unemploy"\}
\end{itemize}
We choose only substrings that are found in the top $\sim 1000$ most frequent unique words, as well as only those we can unambiguously identify with a given subtopic (based on randomly sampling 100 tweets per substring for manual verification). \\ 

In the interest of interpretability, we use a very simple measure to quantify daily average OGE. Let $T^{topic}_C(t)$ be the total number of tweets on day $t$ for city $C$ mentioning the subtopic $topic$, with
$P^{topic}_C(t)$, $M^{topic}_C(t)$ and $N^{topic}_C(t)$ the corresponding subset of these tweets associated with positive, neutral/mixed, and negative sentiment respectively such that $T^{topic}_C(t) = P^{topic}_C(t)+M^{topic}_C(t)+N^{topic}_C(t)$. To find the average polarity of tweets on day $t$ related to a given subtopic $topic$ in a city $C$, which we will call the ``Geolocalized Mean Sentiment" (GMS) regarding \emph{topic}, we assign a score of $+1$ to positive tweets, $0$ to neutral tweets, and $-1$ to negative tweets, and take the average score. Mathematically, the GMS, $G$, is given by  
\begin{align}
\label{eq:GMS}
G^{topic}_C(t) = \frac{P^{topic}_C(t)\times (+1)+M^{topic}_C(t)\times (0)+N^{topic}_C(t)\times (-1)}{T^{topic}_C(t)}
=\frac{P^{topic}_C(t)-N^{topic}_C(t)}{T^{topic}_C(t)}.    
\end{align}
We can see that $G^{topic}_C(t)$ just amounts to the difference in the fraction of the tweets $T^{topic}_C(t)$ that are positive and the fraction that are negative, and is constrained to $[-1,1]$ with $-1$ indicating entirely negative tweets and $+1$ entirely positive tweets. Tweets of neutral and mixed sentiment are accounted for here in that the GMS is diminished in magnitude when they comprise a greater relative fraction of tweets for that day. \\

We define an additional online geolocalized emotion measure based on these GMS values, to quantify the amount of agreement in $G^{topic}_C(t)$ between all pairs of cities $\{C_1,C_2\}$ with respect to all five subtopics of interest simultaneously. First, we construct the GMS vector for each city $C$
\begin{align}
\label{eq:GMS_vector}
\vec{G}_C(t)\equiv \{G^{distancing}_C(t),G^{China}_C(t),G^{Trump}_C(t),G^{economy}_C(t),G^{mask}_C(t)\},    
\end{align}
which takes the form of a vector in $\mathbb{R}^5$. This construction associates each GMS topic with an orthogonal unit axis, which is consistent with our definition of these topics as independent topics of interest. We then define the angle $\theta_{C_1C_2}(t)$ between the vectors $\vec{G}_{C_1}(t)$ and $\vec{G}_{C_2}(t)$ (measured in degrees) as
\begin{align}
\label{eq:angle}
\theta_{C_1C_2}(t) = \cos^{-1}\left(\frac{\vec{G}_{C_1}(t)\cdot \vec{G}_{C_2}(t)}{\vert\vert \vec{G}_{C_1}(t)\vert\vert\; \vert\vert \vec{G}_{C_2}(t)\vert\vert}\right),    
\end{align}
which will be $0^\circ$ when the cities $C_1$ and $C_2$ have collinear GMS vectors, $90^\circ$ when cities $C_1$ and $C_2$ have completely orthogonal GMS vectors, and $180^\circ$ when cities $C_1$ and $C_2$ have anti-parallel GMS vectors. Finally, we compute the ``GMS coherence" $\phi(t)$ as 
\begin{align}
\label{eq:coherence}
\phi(t) = \frac{2}{n_C(n_C-1)}\sum_{C_1\neq C_2}\theta_{C_1C_2}(t),   \end{align}
where $n_C=49$ is the number of cities. This simply gives the ``average angle" between the GMS vectors at time $t$, and is used as a proxy for the general agreement in GMS values over all cities at each time period. In particular, high values of Eq. \ref{eq:coherence} indicate low coherence in the GMS vectors across cities (as the average angle between them is high), and values near $0^\circ$ indicate high coherence, as the GMS vectors are all oriented similarly.\\

We note here that GMS (Eq. \ref{eq:GMS}) and GMS coherence (Eq. \ref{eq:coherence}) are used as simple metrics to capture online geolocalized emotion, but numerous similar constructions for OGE measures are possible. Assuming we've decided on a framework to classify tweet sentiment (a difficult problem in its own right that we will not address here \cite{lima2015polarity}), the GMS could easily be constructed using the average of weighted sentiment scores output by this algorithm, rather than the more coarse approach assigning only values in $\{1,0,-1\}$. However, for this study we choose the GMS measure in Eq. \ref{eq:GMS} so as to not attempt to assign physical significance to the confidence scores output by the sentiment classifcation API, as these are constructed using an unknown proprietary method. Additionally, we should note that the coherence measure in Eq. \ref{eq:coherence} could be adapted to account for correlations between the subtopics by not treating them as orthogonal axes. For this alteration we could simply apply a coordinate transformation to the inner product in Eq. \ref{eq:angle}, making it the inverse covariance matrix between the bias vectors, and transform the norms in the denominator accordingly. (This procedure is more formally called transforming to ``Mahalanobis space" \cite{de2000mahalanobis}.) However, here we again opt for the simpler, more interpretable option of treating the subtopics as orthogonal unit axes, and stress that from experimentation the Mahalanobis transformation gives very qualitatively similar results for the coherence in Fig. 1C.

\subsection{Time series analyses}

We use time series analysis---in particular intervention, correlation, and Granger causality analysis---to examine the relationships of GMS with policy development, offline mobility and infection data. For all time series analyses, we preprocess the series so that they are stationary to remove temporal trends. To do this, for any pair of series ${x_t,y_t}$ that are being compared, we use the following procedure
\begin{enumerate}
    \item Perform an Augmented Dickey-Fuller (ADF) unit root test to check for stationarity of both series
    \item If one or both series fail to reject the null-hypothesis (that there is a unit root in the series) at the $0.05$ significance level, transform both series by taking differences $\{x_t,y_t\}\to \{x_t-x_{t-1},y_t-y_{t-1}\}$
    \item Repeat 1 and 2 until null-hypothesis is rejected at the $0.05$ significance level
\end{enumerate}
Most series pairs only needed to be differenced once to satisfy these criteria, and in the worst case had to be differenced three times.  For all time series analysis involving case or death data, both series are truncated to start when cases or deaths in the associated city become non-zero (which is necessary to pass the stationarity tests anyway). We also note that the application of variance-stabilizing transformations (in particular square-roots and logarithms) did not in general reduce the order of integration for the time series, and so we do not apply these to the data.\\

To assess whether or not a each policy event had a significant impact on the GMS $G^{(topic)}_C(t)$, we use an ARMAX (Auto Regressive Moving Average with with eXplanatory variables) model, which can account for effects from the lagged dependent variable $G^{(topic)}_C(t)$ as well as exogenous categorical (binary) inputs $z_{kt}$ (policy events) \cite{pankratz2012forecasting}. In the ARMAX($p$,$q$) process, the dependent GMS variable is modeled as 
\begin{align}
\label{eq:ARMAX}
G^{(topic)}_C(t) = \sum_{i=1}^{p}\alpha_iG^{(topic)}_C(t-i)-\sum_{i=1}^{q}\gamma_i\epsilon_{t-i}+ \sum_{k=1}^{4}\beta_k z_{kt} +\epsilon_t,
\end{align}
where $p$ is the number of autoregressive terms, $q$ is the number of moving average terms, $k$ indexes the policy events, and $\epsilon_t$ is Gaussian white noise. For the declaration of national emergency ($k=1$ or $k=\text{national emergency declaration}$), extension of social distancing guidelines ($k=2$ or $k=\text{extension of distancing guidelines}$), and issuing of state reopening guidelines ($k=3$ or $k=\text{reopening guidelines announcement}$), we set $z_{kt}=1$ only on the date $t$ of the announcement, and $0$ for all other $t$, while for the local shelter in place orders ($k=4$ or $k=\text{local shelter in place order}$), we set $z_{kt}=1$ for the entire duration of the shelter in place order for each city and $z_{kt}=0$ otherwise. The ARMAX model amounts to a special case of the more general ``transfer function" (or ``dynamic regression") models \cite{box2015time}, tools commonly used in econometric intervention analyses, which can be seen through rewriting Eq. \ref{eq:ARMAX} in the following form
\begin{align}
G^{(topic)}_C(t) = \frac{1}{\alpha(\Delta)}\sum_{k=1}^{4}\beta_kz_{kt}+\frac{\gamma(\Delta)}{\alpha(\Delta)}\epsilon_t,    
\end{align}
where $\Delta$ is the differencing (or ``backshift") operator that transforms $x_t\to x_{t-1}$, and $\alpha(\Delta) = 1-\sum_{i=1}^{p}\alpha_i\Delta^i$ and $\gamma(\Delta) = 1-\sum_{i=1}^{q}\gamma_i\Delta^i$. \\

For each intervention analysis, we scan over the range of lags $(p,q)\in [0,7]\times [0,7]$ and pick the pair $(p,q)$ with the lowest Bayesian Information Criterion (BIC), a model selection diagnostic based on the fit likelihood with a penalty for less parsimonious models \cite{box2015time}. Additionally, autocorrelation and partial autocorrelation functions of randomly sampled series were visually examined to verify the model fits, to ensure that significant autocorrelation lags in the GMS variables were properly accounted for in the ARMAX model. Using the ARMAX($p$,$q$) process, we are able to see the nature and impact of a policy event $z_{kt}$ on $G^{(topic)}_C(t)$, after accounting for correlations from lagged values of this GMS variable and moving average terms, by analyzing the sign and statistical significance of the maximum likelihood estimates $\hat\beta_k$ inferred through this model. In particular, we look at the standardized $z$-score $Z^{(topic)}_k$ associated with each estimate $\hat\beta_k$ for the ARMAX model with dependent variable $G^{(topic)}_{C}(t)$
\begin{align}
\label{eq:zscore}  
Z^{(topic)}_k = \frac{\hat\beta_k}{\sigma_{\beta_k}},
\end{align}
where $\sigma_{\beta_k}$ is the standard error of the estimate $\hat\beta_k$. Using $Z^{(topic)}_{k}$ removes allows us to assess both the sign and statistical significance of $\beta_k$ in a scale-independent manner. Similar to the pairwise analyses, the $G^{(topic)}_C(t)$ series were stationarized through differencing once (all that was needed to reject the ADF null at the $0.05$ significance level for all cities) prior to intervention analysis.  \\

For the correlation analyses in Fig. 1B and 3A, we perform the differencing procedure discussed at the beginning of this section to each pair of variables, and then compute their associated Pearson correlation coefficient. We also implement Granger (non-)causality testing in Fig. 3B to determine whether a given GMS variable, $G^{(topic)}_C(t)$, is theoretically able to provide additional statistically significant information for predicting the future values of each mobility variable, $M$, accounting for past values of $M$ \cite{granger2014forecasting}. First, the optimal lag $p$ for the univariate autoregression of $M$ is determined by fitting
\begin{align}
M_t = \sum_{i=1}^{p}\alpha_iM(t-i)+\epsilon_t    
\end{align}
(where again $\epsilon_t$ is a white noise process, and $\alpha$ are the autoregression coefficients) at a range of $p$'s and selecting the best fitting model using the associated BIC. Then, we add in the lagged values of $G^{(topic)}_C(t)$ and fit 
\begin{align}
\label{eq:granger}
M_t = \sum_{i=1}^{p}\alpha_iM(t-i)+\sum_{i=1}^{p}\omega_iG^{(topic)}_C(t-i)+\epsilon_t.     
\end{align}
Finally, we reject the null hypothesis that $G^{(topic)}_C$ does not Granger-cause $M$ if any of the $\omega_i$ are determined to be significantly different than $0$ through chi-squared testing \cite{bessler1984note}. GMS and mobility variables are differenced to the same level prior to testing, using the procedure outlined earlier.

\section{Results}
\subsection{Universal trends in online geolocalized emotional responses across US cities}
As a first step in understanding the behavior of online geolocalized emotion (OGE) across the US during the early stages of the COVID epidemic in the US, the daily geolocalized mean sentiment (GMS, Eq. \ref{eq:GMS}) with respect to the five subtopics discussed in Section IIB is plotted across time in Fig. 1A at the national level---computed based on all daily subtopic relevant tweets in the dataset, irrespective of location---and for five geographically dispersed example cities. We also show the dates of the policy events identified in Section IIA for reference, and values are plotted using a weekly moving average to smooth out fluctuations for easier visualization. The first pattern we can observe is the strong similarity between the trends in the national-level GMS and the GMS in the five example cities shown, for all subtopics. This suggests that, at the city level, there is little heterogeneity in OGE trends geographically, in contrast to policy response across local governments for which there has been a high level of heterogeneity \cite{haffajee2020thinking,goolsbee2020COVID}. These universal average trends in OGE across cities may also point to a greater level of commonality in local public collective emotional responses to COVID than is perhaps suggested by the variation in local mobility, which is known to be associated with happiness levels \cite{frank2013happiness}. \\

Looking at the trends in Fig. 1A in more detail, we can see in general that GMS is negative for all subtopics, and there is a similar ordering in the GMS values for the subtopics and overall GMS over time, with ``distancing" typically garnering the most positive GMS and ``TA" typically garnering the most negative GMS. We also see relative stability in ``TA" and ``China" GMS values, while ``distancing", ``economy", ``mask", and overall GMS values tend to increase over the period studied. We can also observe that there are relatively strong fluctuations in the trends for many of the GMS series near the date of President Trump's extension of social distancing guidelines, particularly in ``distancing", which even reaches above zero during this period for all examples shown. This reflects a general sense of positivity about social distancing-related behaviors surrounding this announcement, perhaps indicating people's commitment to, or resilience regarding, continued distancing. \\

To compliment our qualitative visual analysis in Fig. 1A., in Fig. 1B and 1C we investigate more quantitatively whether or not there is a strong correlation in the GMS values across these cities. Shown in Fig. 1B are the distributions of Pearson correlation coefficients between city-level GMS values across all pairs of cities, for overall GMS and all subtopics of GMS. Series were stationarized through differencing once prior to analysis, and so the correlations we see are actually between day-to-day changes in GMS values. Also shown in Fig. 1B below each boxplot is the corresponding percentage of all Pearson correlations that were statistically significant at the $0.05$ level for that subtopic. We can see that the Pearson correlations across all city pairs are relatively high for overall GMS and all GMS subtopics, indicating that not only the temporal trends are similar between these GMS variables, but the daily fluctuations are also highly correlated. This is reflected in the high percentage of significant correlations as well.\\

In Fig. 1C we assess a different dimension to this homogeneity in GMS, aggregating GMS with respect to all subtopics (excluding overall GMS) into temporal vectors and looking at the similarity in these vectors over time through the coherence measure in Eq. \ref{eq:coherence}. We also plot a moving average, this time a three-day moving average, to more easily visualize trends. We can see that the GMS coherence over time is pretty stable, fluctuating between $\approx 13^\circ$ and $\approx 24^\circ$ (relative to a maximum value of $180^{\circ}$), and maintaining relatively low values, indicating high similarity in the GMS vectors over time, consistent with the findings in the other two panels. However, we also observe some disturbances in the pattern occurring around federal policy event dates. More specifically, we see an increase in GMS vector similarity (through a declining coherence measure) near the extension of social distancing guidelines, and a transition in the trend around the state reopening guidelines announcement from relatively unchanging to increasing. Noting the generally high similarity in the OGE dynamics across cities seen in Fig. 1A and Fig. 1B, as well as the low values of Eq. \ref{eq:coherence} in Fig. 1C, these disturbances provide initial evidence that we still see heterogeneity in city-level OGE, but it manifests itself in the direct influence of policy events, which is a much more subtle factor to address. We perform the analysis necessary to address these effects in the next section.

\begin{figure}
    \centering
    \includegraphics[width=1\textwidth]{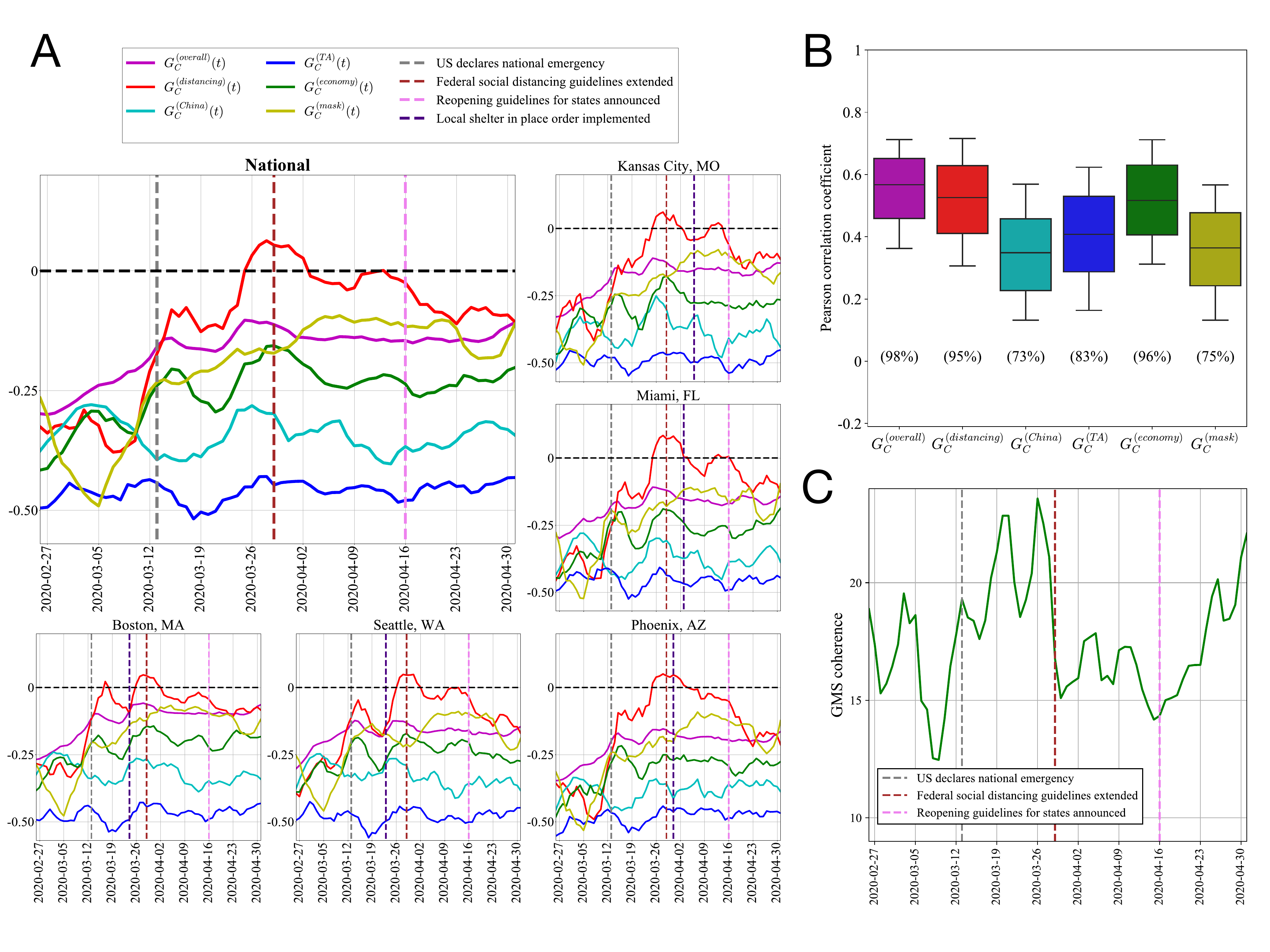}
    \caption{\textbf{Universal trends in online geolocalized emotion across US cities during the early stages of the epidemic.} \textbf{(A)} Temporal trends for all GMS values considered (Eq. \ref{eq:GMS}) over the time period studied (solid lines), for the entire dataset as well as for a selection of geographically dispersed cities, showing strikingly uniform temporal trends. Key policy events identified during this time period (dashed vertical lines) are shown for reference, and 7-day moving averages are displayed for the time series for clearer visualization. \textbf{(B)} Distribution of Pearson correlation coefficients between all pairs of (differenced) city-level GMS time series, for each GMS type, indicating high correlation in day-to-day GMS values across cities. Whiskers denote the 10th and 90th percentiles, and below each boxplot we display the percentage of all Pearson correlations for the corresponding GMS subtopic that were statistically significant at the $0.05$ level. \textbf{(C)} GMS coherence (Eq. \ref{eq:coherence}) over the time period studied (solid green line), indicating low variability in cities' GMS vectors (Eq. \ref{eq:GMS_vector}) across time, with particularly consistent GMS values between the extension of distancing guidelines and the announcement of reopening procedures.}
    \label{fig:fig1}
\end{figure}

\subsection{Sensitivity to federal and local policies}

To assess the extent to which each major policy event (detailed in Section IIA) has an effect on the dynamics of online geolocalized emotion in each city, we perform the intervention analysis discussed in Section IIC for each subtopic GMS $G_C^{(topic)}$ and each city $C$, extracting the effect sizes $Z_k^{(topic)}$ in Eq. \ref{eq:zscore}. In Fig. 2A we show the distribution of these intervention effect sizes $Z_k^{(topic)}$ regarding each subtopic for all cities across the four events identified for analysis. We observe generally strong effect sizes for the declaration of national emergency and local shelter in place orders, while the extension of distancing guidelines and announcement of state reopening guidelines have relatively weak values of $Z_{k}^{(topic)}$. The declaration of national emergency appears to have a very mixed effect on GMS values, with the distributions for $Z^{(China)}_{1}$ and $Z^{(TA)}_{1}$ displaying a strong tendency towards negative values, and the other variables showing a tendency towards positive values. Around $50\%$ of cities have ``China" and ``TA" subtopic GMS dynamics that are negatively affected by the declaration of national emergency to a statistically significant extent, indicating that these two topics were associated with a high level of negative sentiment as a result of the declaration. This is consistent with the high level of anti-Chinese sentiment observed during the early stages of the epidemic \cite{devakumar2020racism}, and expressions of anger on social media towards both US leadership and China \cite{li2020analyzing}, although here we gain a more nuanced understanding of the effect a specific event has on these responses at a local level. We can also see generally negative intervention effect sizes connected with the implementation of local shelter in place orders, particularly on overall, ``distancing" and ``TA" GMS values, perhaps reflecting the anger and frustration associated with quarantining \cite{brooks2020psychological}. We compliment this illustrative analysis of intervention effect sizes in Fig. 2B with a visualization displaying the sensitivity of each city studied to the policy events, as measured by the total number of policies by which the city's overall GMS values were affected to a statistically significant extent (at the $0.05$ level). Here we see a moderate geographic trend, with cities in the southeastern US and Texas having generally more significant policy responses, and cities in the northeastern and southwestern US having generally fewer significant policy responses. However, we still see variability within each region, and so these responses are not necessarily well localized in space. \\

The sensitivity of cities to policy events is further investigated in Fig. 2C, where we plot the intervention effect sizes $Z^{(topic)}_K$ for different policy events $k$ for the same city on each axis, with the three panels showing different subtopics \emph{topic}. We perform three OLS linear regressions, one for each of the three pairs of variables, and determine through low autocorrelation of approximately normally distributed residuals as well as low $p$-values in all cases that these linear fits are appropriate models for the data. We observe in the top panel of Fig. 2C that cities responding more negatively about distancing after the national emergency declaration also tend to respond more negatively about distancing after their local shelter in place orders, and likewise for cities responding positively. We also see the same trend for the GMS responses relevant to the economy in the middle panel. The bottom panel shows that when comparing responses to the reopening guidelines announcement and responses to local shelter in place orders, we actually see the opposite effect, at least regarding ``TA" biases. The negative correlation we see in this panel may reflect the different natures of the local shelter in place orders and the reopening guidelines announcement: for cities that respond negatively to local shelter in place orders, the announcement of reopening guidelines may be seen as a statement of optimism. On the other hand, for cities that respond positively to local shelter in place orders, the announcement of reopening guidelines may seem premature. However, making these determinations conclusively requires a more contextualized analysis with the aggregation of data from different sources. The results in Fig. 2 altogether indicate that some cities tend to be more sensitive to federal and local policy in their OGE dyanmics than others, and comparison with the results in Fig. 1 suggests that the heterogeneity in city-level OGE dynamics is better reflected by cities' GMS responses to policy events rather than the overall observed trends and day-to-day correlations in GMS. Keeping in mind these observations about the manifestation of OGE at the city-level, we transition in the next section to analyzing its connection with local offline factors such as epidemic indicators and human mobility. 

\begin{figure}
    \centering
    \includegraphics[width=1\textwidth]{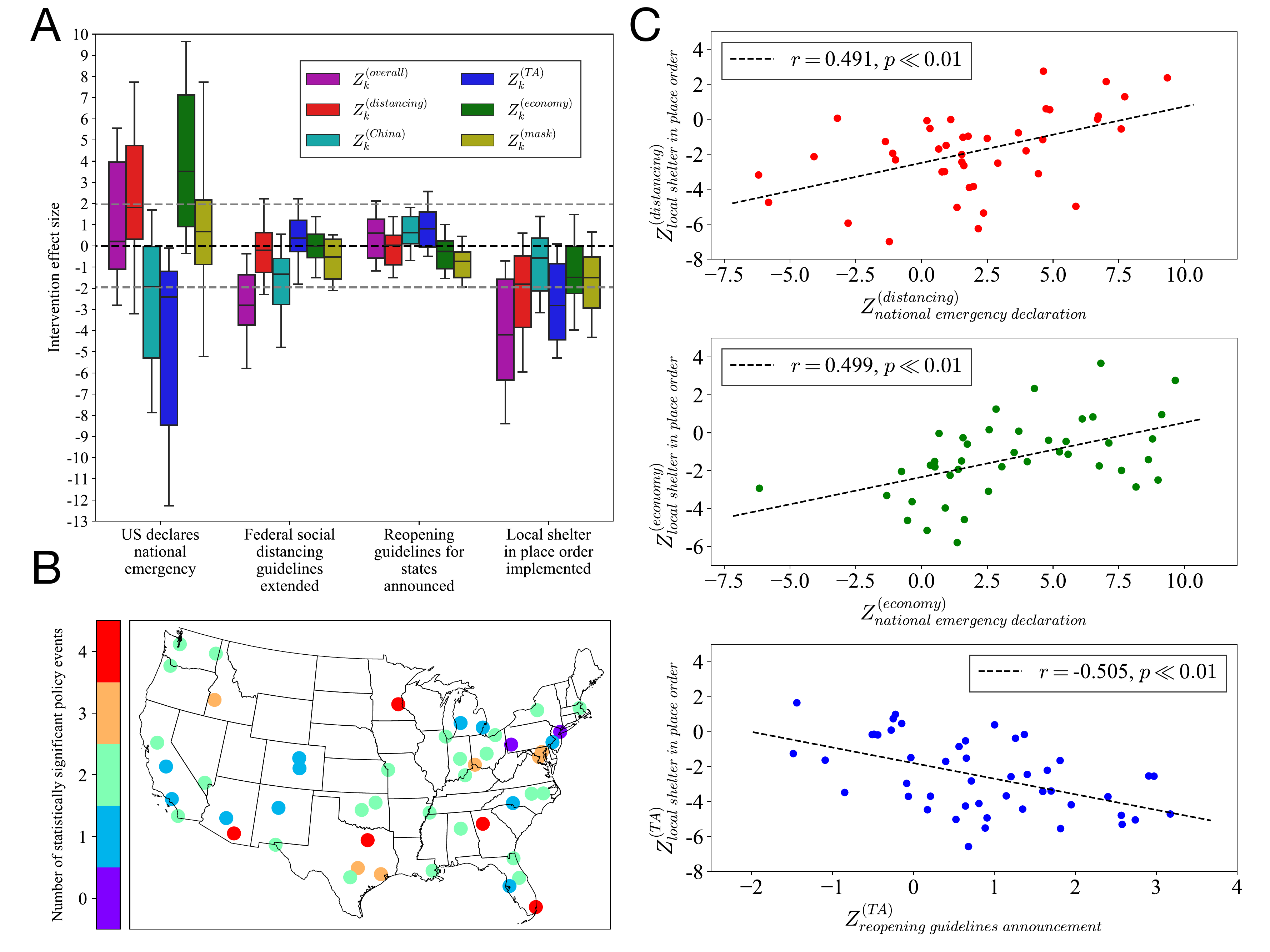}
    \caption{\textbf{Effects of major federal and local policies on online geolocalized emotion.} \textbf{(A)} Policy effect sizes $Z^{(topic)}_k$ (Eq. \ref{eq:zscore}) for all GMS values \emph{topic} and policy events $k$, displaying high variation across cities in response to the national emergency declaration and local shelter in place orders, and lower sensitivity to the announcement of federal guidelines for social distancing extension and state reopening. Also shown are gray horizontal lines indicating the positions at which effect sizes are statistically significant at the $0.05$ level with respect to a standard normal distribution. \textbf{(B)} Number of statistically significant responses in overall GMS $Z^{(overall)}_k$ (at the $0.05$ level) to events $k\in [0,4]$, for all cities, indicating the general sensitivity of cities to the policies enacted during the period of study. \textbf{(C)} Effect sizes for various policy events on $B^{(distancing)}$ (top), $B^{(economy)}$ (middle), and $B^{(TA)}$ (bottom), showing high Pearson correlations $r$ between GMS responses regarding both displayed policy events for each topic. OLS regressions were performed too det Extreme outliers with $\abs{Z^{(topic)}_k}>>10$ are omitted from the OLS regressions, and the associated $p$-values are displayed alongside the Pearson correlation coefficients $r$.}
    \label{fig:fig2}
\end{figure}

\subsection{High associations between online emotional responses and offline mobility}

As a final investigation into the dynamics of online geolocalized emotion in cities, we look at its association with epidemic indicators (daily cases and deaths) and mobility measures (relative walking and driving volume). Further explanation of the epidemic and mobility datasets integrated into our analysis can be found in Section IIA. In Fig. 3A, we plot the Pearson correlation coefficient for all pairs of offline and GMS variables to determine the strength and nature of the unlagged temporal correlation between these quantities within each city. Each pair of variables was differenced until both were stationary by using the procedure discussed in Section IIC, which is crucial for eliminating the confounding temporal trends in all the variables studied. It is reasonable to guess that epidemic indicators may have instantaneous daily correlations with OGE: the abundance of online publicly available data and constant national and local media coverage of case and death statistics results in high, instant exposure to epidemic updates, the psychological effects from which have been discussed at length in current research \cite{dong2020letter,xiong2020impact}. However, we can see from Fig. 3A that epidemic statistics actually have very little correlation with GMS values. Only two pairs of variables involving epidemic indicators have statistically significant correlations in more than ten cities, while nine pairs involving mobility measures do. We can also see that among these generally low correlations, national daily cases and deaths have significant correlations with GMS in substantially more cities than local daily cases and deaths. In general, the significant correlations between case/death data and GMS values tend to be negative, indicating that emotional responses have a greater negative tendency as epidemic indicators grow more rapidly.\\

As opposed to epidemic indicators, we find that mobility measures are consistently highly correlated with many of the GMS subtopic measures. In particular, we see consistently strong and statistically significant negative correlations between mobility and $G^{(distancing)}_C$ as well as $G^{(economy)}_{C}$ in many cities. We also see strong correlations between mobility measures and $G^{(overall)}_C$ as well as $G^{(China)}_C$, though in fewer cities, and $G^{(TA)}$ and $G^{(mask)}$ appear to have much weaker associations with mobility measures. An interesting aspect of these correlations is that they are actually different in sign among various GMS subtopics: ``distancing", ``economy", and overall GMS values tend to have negative correlations with mobility measures, while ``China", ``mask", and ``TA" GMS values tend to have positive correlations with mobility measures. Based on these mixed relationships between GMS subtopics and offline mobility, the impact of mobility on OGE regarding each subtopic individually is unclear, though we do know that there is a consistent statistical association between these quantities. The underlying psychological reasons for these connections between OGE and mobility can be investigated by future studies. However, for practical risk management, if we can use this online geolocalized emotion to predict future mobility patterns, this can aid in effective intervention plans.  We thus look at a more general formulation of statistical association in Fig. 3B, assessing whether or not past GMS values can theoretically provide statistically relevant information about future mobility.\\

In Fig. 3B we show the optimal Granger causality lag for all pairs \{GMS variable, mobility variable\} that have a statistically significant Granger-causal relationship (details given in Section IIC). Interpreting Granger causality as an indicator of theoretical predictability, we can see that GMS can consistently be used to aid in the prediction of future mobility values for most cities, and that frequently this prediction is possible at lags of a week or greater. We also note that the cities with the largest populations---in particular New York, Los Angeles, and Chicago---have significant Granger-causality across nearly all pairs of variables, with longer lag times that tend towards two weeks due to long-range autocorrelations in the mobility values in these areas. Investigating the causes of this peculiar pattern, however, is outside the scope of this work. These results, along the correlations seen in Fig. 1A, suggest that there is a high statistical association between OGE and mobility at the daily level, and that the former can be effectively used to aid prediction of the latter with substantial foresight.

\begin{figure}
    \centering
    \includegraphics[width=1\textwidth]{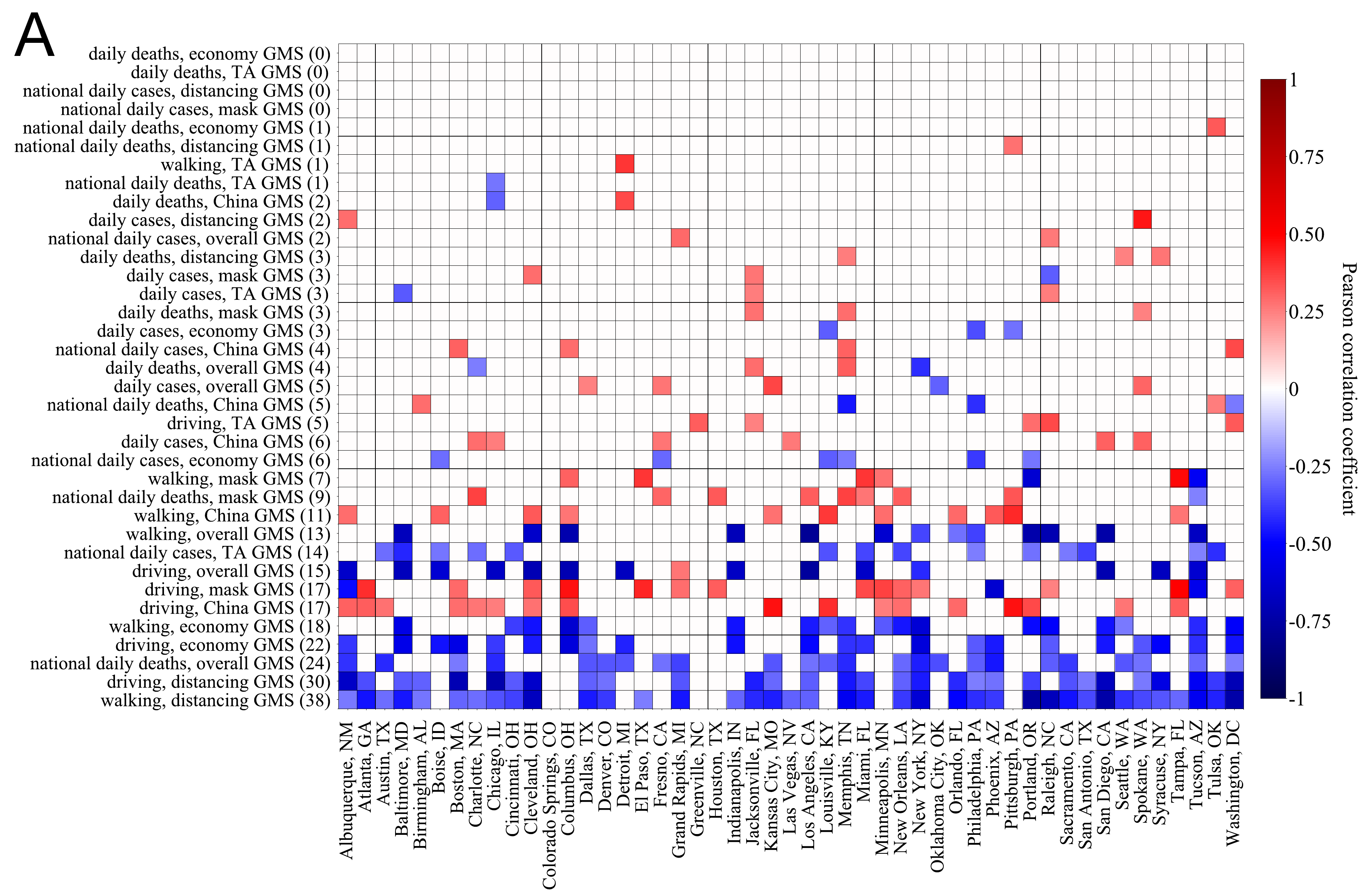}\\
    \includegraphics[width=1\textwidth]{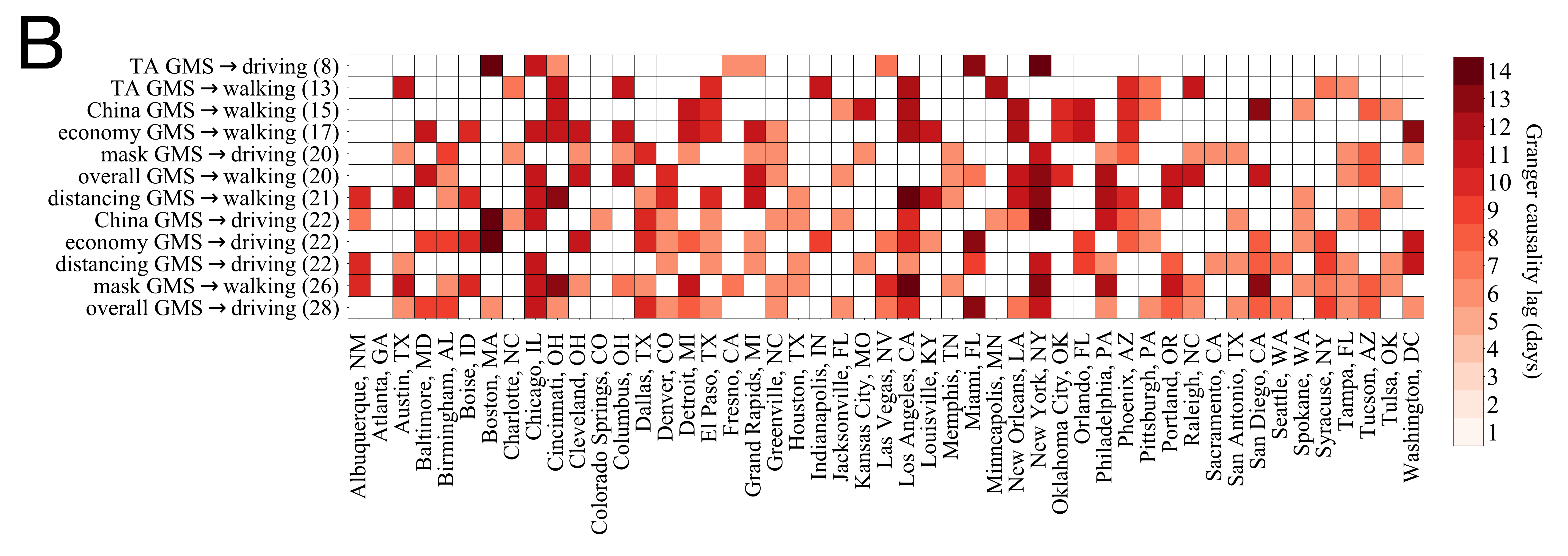}
    \caption{\textbf{Association between online geolocalized emotion and offline factors.} \textbf{(A)} Pearson correlations between (stationarized) GMS values and offline indicators, showing high correlation between GMS and mobility measures, but weaker correlations with epidemic measures. Only correlations that are statistically significant at the $0.05$ level are shaded, and rows are ordered top to bottom by the number of cities with a significant correlation between the corresponding measures, which is shown in parenthesis alongside each pair of variables. \textbf{(B)} Inferred statistically significant Granger causality lags (Eq. \ref{eq:granger}) for mobility measures with lagged GMS variables, indicating that prediction of future mobility in cities is consistently aided through the information contained in the GMS values in most cities. Again, only lags for causality tests significant at the $0.05$ level are shaded, and rows are ordered by number of cities with statistically significant causalities, which is labelled in parenthesis alongside variable pairs.}
    \label{fig:fig3}
\end{figure}

\section{Conclusions and Future Work}
 In this study, we examine online geolocalized emotional (OGE) responses towards COVID and five related subtopics across 49 US cities from Feb 26th to May 1st 2020 using a dataset of around 13 million tweets with geolocation attributes. We assess the temporal dynamics of OGE in these cities through a few sentiment-derived measures, as well as analyze the associations of OGE dynamics with critical COVID-relevant policy events, offline mobility, and epidemic measures. The key findings of this project related to our original research questions are: 1) There is a universal temporal trend in OGE across US cities, with high day-to-day correlations and consistent relative negativity in sentiment across the COVID subtopics; 2) OGE across cities is sensitive to major federal policy announcements and local shelter in place orders, and some cities are much more consistently sensitive to policy events than others; 3) OGE is highly correlated with mobility but not with epidemic measures, and OGE has a high predictive capability for future mobility. The findings of this study help us to understand the city-level manifestation of public online emotional responses during the COVID crisis in the US, and how this online collective emotion connects with offline factors such as policy, epidemic measures, and human mobility. \\

There is a plethora of possible future work extending the ideas presented in this study. One clear avenue for future work is the extension of the timeframe studied to incorporate data up to the present day, and the application of these methods to OGE dynamics in cities worldwide. Another important extension is to incorporate a more refined set of sentiment classifications---for example including classifications for fear or anger as subsets of negative sentiment---and constructing new OGE measures based on these categories. Additionally, our framework can be adapted to examine other important aspects of public behavioral responses (such as purchasing behavior) or demographic factors (such as socioeconomic status), and how these connect with online geolocalized emotion during crises. Finally, the practical application of predicting future values of mobility using OGE is a critical avenue for study that builds off of this project, which can be used in conjunction with existing studies assessing the impact of human mobility on epidemic spread to make informed policy decisions \cite{kraemer2020effect,aguilar2020impact,badr2020association}.


\begin{thebibliography}{10}
\expandafter\ifx\csname url\endcsname\relax
  \def\url#1{\texttt{#1}}\fi
\expandafter\ifx\csname urlprefix\endcsname\relax\def\urlprefix{URL }\fi

\bibitem{gao2011harnessing}
H.~Gao, G.~Barbier, and R.~Goolsby, Harnessing the crowdsourcing power of
  social media for disaster relief. \textit{IEEE Intelligent Systems}
  \textbf{26}, 10 (2011).

\bibitem{neubaum2014psychosocial}
G.~Neubaum, L.~R{\"o}sner, A.~M. Rosenthal-von~der P{\"u}tten, and N.~C.
  Kr{\"a}mer, Psychosocial functions of social media usage in a disaster
  situation: A multi-methodological approach. \textit{Computers in Human
  Behavior} \textbf{34}, 28 (2014).

\bibitem{li2017reasoning}
X.~Li, Z.~Wang, C.~Gao, and L.~Shi, Reasoning human emotional responses from
  large-scale social and public media. \textit{Applied Mathematics and
  Computation} \textbf{310}, 182 (2017).

\bibitem{gao2020mental}
J.~Gao, P.~Zheng, Y.~Jia, H.~Chen, Y.~Mao, S.~Chen, Y.~Wang, H.~Fu, and J.~Dai,
  Mental health problems and social media exposure during COVID-19 outbreak.
  \textit{PLoS One} \textbf{15}, 0231924 (2020).

\bibitem{roy2020study}
D.~Roy, S.~Tripathy, S.~K. Kar, N.~Sharma, S.~K. Verma, and V.~Kaushal, Study
  of knowledge, attitude, anxiety \& perceived mental healthcare need in Indian
  population during COVID-19 pandemic. \textit{Asian Journal of Psychiatry} \textbf{51}, 102083 (2020).

\bibitem{zhang2020impact}
Y.~Zhang and Z.~F. Ma, Impact of the COVID-19 pandemic on mental health and
  quality of life among local residents in Liaoning province, China: A
  cross-sectional study. \textit{International Journal of Environmental
  Research and Public Health} \textbf{17}, 2381 (2020).

\bibitem{barkur2020sentiment}
G.~Barkur and G.~B.~K. Vibha, Sentiment analysis of nationwide lockdown due to
  COVID 19 outbreak: Evidence from India. \textit{Asian Journal of Psychiatry} \textbf{51}, 102089 (2020).

\bibitem{mckay2020anxiety}
D.~McKay, H.~Yang, J.~Elhai, and G.~Asmundson, Anxiety regarding contracting
  COVID-19 related to interoceptive anxiety sensations: The moderating role of
  disgust propensity and sensitivity. \textit{Journal of Anxiety Disorders} \textbf{73}, 102233 (2020).

\bibitem{zhang2020monitoring}
Y.~Zhang, H.~Lyu, Y.~Liu, X.~Zhang, Y.~Wang, and J.~Luo, Monitoring depression
  trend on Twitter during the COVID-19 pandemic. \textit{arXiv preprint
  arXiv:2007.00228} (2020).

\bibitem{gupta2020COVID}
R.~K. Gupta, A.~Vishwanath, and Y.~Yang, COVID-19 Twitter dataset with latent
  topics, sentiments and emotions attributes. \textit{arXiv preprint
  arXiv:2007.06954} (2020).

\bibitem{sun2020early}
K.~Sun, J.~Chen, and C.~Viboud, Early epidemiological analysis of the
  coronavirus disease 2019 outbreak based on crowdsourced data: a
  population-level observational study. \textit{The Lancet Digital Health}
  (2020).

\bibitem{pulido2020COVID}
C.~M. Pulido, B.~Villarejo-Carballido, G.~Redondo-Sama, and A.~G{\'o}mez,
  COVID-19 infodemic: More retweets for science-based information on
  coronavirus than for false information. \textit{International Sociology} 
  \textbf{35}, 0268580920914755 (2020).

\bibitem{tao2020nature}
Z.-Y. Tao, G.~Chu, C.~McGrath, F.~Hua, Y.~Y. Leung, W.-F. Yang, and Y.-X. Su,
  Nature and diffusion of COVID-19--related oral health information on Chinese
  social media: analysis of tweets on Weibo. \textit{Journal of Medical
  Internet Research} \textbf{22}, 19981 (2020).

\bibitem{thelwall2020retweeting}
M.~Thelwall and S.~Thelwall, Retweeting for COVID-19: Consensus building,
  information sharing, dissent, and lockdown life. \textit{arXiv preprint
  arXiv:2004.02793} (2020).

\bibitem{memon2020characterizing}
S.~A. Memon and K.~M. Carley, Characterizing COVID-19 misinformation
  communities using a novel Twitter dataset. \textit{arXiv preprint
  arXiv:2008.00791} (2020).

\bibitem{sharma2020COVID}
K.~Sharma, S.~Seo, C.~Meng, S.~Rambhatla, and Y.~Liu, COVID-19 on social media:
  Analyzing misinformation in Twitter conversations. \textit{arXiv preprint
  arXiv:2003.12309} (2020).

\bibitem{shen2020using}
C.~Shen, A.~Chen, C.~Luo, J.~Zhang, B.~Feng, and W.~Liao, Using reports of own
  and others' symptoms and diagnosis on social media to predict COVID-19 case
  counts: Observational infoveillance study in mainland China. \textit{Journal
  of Medical Internet Research} \textbf{22} 19421 (2020).

\bibitem{porcher2020social}
S.~Porcher and T.~Renault, Social distancing beliefs and human mobility:
  Evidence from Twitter. \textit{arXiv preprint arXiv:2008.04826} (2020).

\bibitem{huang2020twitter}
X.~Huang, Z.~Li, Y.~Jiang, X.~Li, and D.~Porter, Twitter, human mobility, and
  COVID-19. \textit{arXiv preprint arXiv:2007.01100} (2020).

\bibitem{qazi2020geocov19}
U.~Qazi, M.~Imran, and F.~Ofli, Geocov19: a dataset of hundreds of millions of
  multilingual COVID-19 tweets with location information. \textit{SIGSPATIAL
  Special} \textbf{12}, 6 (2020).

\bibitem{censusMSA}
United States Census Bureau Population Division, \textit{Metropolitan and
Micropolitan Statistical Areas Population Totals and Components of Change:
2010-2019} (2019).

\bibitem{saif2012alleviating}
H.~Saif, Y.~He, and H.~Alani, Alleviating data sparsity for Twitter sentiment
  analysis. In \emph{21st International Conference on the World Wide Web} (CEUR Workshop Proceedings, Lyon, 2012), pp. 2-9. 
  
 
\bibitem{srivastava2017challenges}
R.~Srivastava and M.~Bhatia, Challenges with sentiment analysis of on-line
  micro-texts. \textit{International Journal of Intelligent Systems and
  Applications} \textbf{9}, 31 (2017).

\bibitem{carvalho2020off}
A.~Carvalho and L.~Harris, Off-the-shelf technologies for sentiment analysis of
  social media data: Two empirical studies. In \textit{AMCIS 2020 Proceedings} (Association for Information Systems Library, 2020).

\bibitem{lwin2020global}
M.~O. Lwin, J.~Lu, A.~Sheldenkar, P.~J. Schulz, W.~Shin, R.~Gupta, and Y.~Yang,
  Global sentiments surrounding the COVID-19 pandemic on twitter: Analysis of
  Twitter trends. \textit{JMIR Public Health and Surveillance} \textbf{6},
  19447 (2020).

\bibitem{carter2020making}
D.~P. Carter and P.~J. May, Making sense of the us COVID-19 pandemic response:
  A policy regime perspective. \textit{Administrative Theory \& Praxis} \textbf{42}, 265 (2020).

\bibitem{parodi2020containment}
S.~M. Parodi and V.~X. Liu, From containment to mitigation of COVID-19 in the
  US. \textit{Jama} \textbf{323}, 1441 (2020).

\bibitem{moon2020fighting}
M.~J. Moon, Fighting against COVID-19 with agility, transparency, and
  participation: Wicked policy problems and new governance challenges.
  \textit{Public Administration Review} \textbf{80}, 651 (2020).

\bibitem{lima2015polarity}
A.~C.~E. Lima, L.~N. de~Castro, and J.~M. Corchado, A polarity analysis
  framework for twitter messages. \textit{Applied Mathematics and Computation}
  \textbf{270}, 756 (2015).

\bibitem{de2000mahalanobis}
R.~De~Maesschalck, D.~Jouan-Rimbaud, and D.~L. Massart, The Mahalanobis
  distance. \textit{Chemometrics and Intelligent Laboratory Systems}
  \textbf{50}, 1 (2000).

\bibitem{pankratz2012forecasting}
A.~Pankratz, \textit{Forecasting with Dynamic Regression Models} (John Wiley \&
  Sons, Hoboken, NJ, 1991).

\bibitem{box2015time}
G.~E. Box, G.~M. Jenkins, G.~C. Reinsel, and G.~M. Ljung, \textit{Time Series
  Analysis: Forecasting and Control}, 4th ed. (John Wiley \& Sons, Hoboken, NJ, 2008).

\bibitem{granger2014forecasting}
C.~W.~J. Granger and P.~Newbold, \textit{Forecasting Economic Time Series}, 2nd ed. (Academic Press, San Diego, 1986).

\bibitem{bessler1984note}
D.~A. Bessler and J.~L. Kling, A note on tests of Granger causality.
  \textit{Applied Economics} \textbf{16}, 335 (1984).

\bibitem{haffajee2020thinking}
R.~L. Haffajee and M.~M. Mello, Thinking globally, acting locally—the US
  response to COVID-19. \textit{New England Journal of Medicine}
  \textbf{382}, 75 (2020).

\bibitem{goolsbee2020COVID}
A.~Goolsbee, N.~B. Luo, R.~Nesbitt, and C.~Syverson, COVID-19 lockdown policies
  at the state and local level. \textit{University of Chicago, Becker Friedman
  Institute for Economics Working Paper} (2020).

\bibitem{frank2013happiness}
M.~R. Frank, L.~Mitchell, P.~S. Dodds, and C.~M. Danforth, Happiness and the
  patterns of life: A study of geolocated tweets. \textit{Scientific Reports}
  \textbf{3}, 1--9 (2013).

\bibitem{devakumar2020racism}
D.~Devakumar, G.~Shannon, S.~S. Bhopal, and I.~Abubakar, Racism and
  discrimination in COVID-19 responses. \textit{The Lancet}
  \textbf{395}, 1194 (2020).

\bibitem{li2020analyzing}
X.~Li, M.~Zhou, J.~Wu, A.~Yuan, F.~Wu, and J.~Li, Analyzing COVID-19 on online
  social media: Trends, sentiments and emotions. \textit{arXiv preprint
  arXiv:2005.14464} (2020).

\bibitem{brooks2020psychological}
S.~K. Brooks, R.~K. Webster, L.~E. Smith, L.~Woodland, S.~Wessely,
  N.~Greenberg, and G.~J. Rubin, The psychological impact of quarantine and how
  to reduce it: Rapid review of the evidence. \textit{The Lancet} \textbf{395}, 912 (2020).

\bibitem{dong2020letter}
M.~Dong and J.~Zheng, Letter to the editor: Headline stress disorder caused by
  Netnews during the outbreak of COVID-19. \textit{Health Expectations: An
  International Journal of Public Participation in Health Care and Health
  Policy} \textbf{23}, 259 (2020).

\bibitem{xiong2020impact}
J.~Xiong, O.~Lipsitz, F.~Nasri, L.~M. Lui, H.~Gill, L.~Phan, D.~Chen-Li,
  M.~Iacobucci, R.~Ho, A.~Majeed, \textit{et~al.}, Impact of COVID-19 pandemic
  on mental health in the general population: A systematic review.
  \textit{Journal of Affective Disorders} \textbf{277}, 55 (2020).

\bibitem{kraemer2020effect}
M.~U. Kraemer, C.-H. Yang, B.~Gutierrez, C.-H. Wu, B.~Klein, D.~M. Pigott,
  L.~Du~Plessis, N.~R. Faria, R.~Li, W.~P. Hanage, \textit{et~al.}, The effect
  of human mobility and control measures on the COVID-19 epidemic in China.
  \textit{Science} \textbf{368}, 493 (2020).

\bibitem{aguilar2020impact}
J.~Aguilar, A.~Bassolas, G.~Ghoshal, S.~Hazarie, A.~Kirkley, M.~Mazzoli,
  S.~Meloni, S.~Mimar, V.~Nicosia, J.~J. Ramasco, \textit{et~al.}, Impact of
  urban structure on COVID-19 spread. \textit{arXiv preprint arXiv:2007.15367} (2020).

\bibitem{badr2020association}
H.~S. Badr, H.~Du, M.~Marshall, E.~Dong, M.~M. Squire, and L.~M. Gardner,
  Association between mobility patterns and COVID-19 transmission in the USA: A
  mathematical modelling study. \textit{The Lancet Infectious Diseases} (2020).

\end{thebibliography}

\end{document}